\documentclass[aps,twocolumn,superscriptaddress,nofootinbib]{revtex4}
\usepackage{exscale}
\usepackage{graphicx}
\usepackage{amsmath}
\usepackage{latexsym}
\usepackage{amsfonts}
\usepackage{amssymb}

\DeclareMathOperator{\Tr}{tr}

\DeclareMathOperator{\vardelta}{\delta}

\DeclareMathOperator{\range}{range}
\newcommand{\field}{\mathbb C}

\newcommand{\calhanc}{ {\cal H}_A \otimes {\cal H}_{Aa} \otimes {\cal
    H}_B \otimes {\cal H}_{Ba}}
                                                                               
\newtheorem{fact}{Fact}
\newtheorem{result}{Result}
\newcommand{\ket}[1]{ | \, #1  \rangle}

\DeclareMathOperator{\eins}{ {\bf 1}}

\begin{document} 

\title{Unitarity as preservation of entropy and 
entanglement in quantum systems}

\author{Florian Hulpke}
\affiliation{Institut f\"ur Theoretische Physik, Universit\"at
  Hannover, D-30167 Hannover, Germany}
\author{Uffe V. Poulsen}
\affiliation{Department of Physics and Astronomy, University of
  Aarhus, DK-8000 Aarhus C., Denmark}
\author{Anna Sanpera}
\altaffiliation[Also at:]{ Instituci\'o Catalana de Recerca i Estudis
  Avan{\c c}ats.}
\affiliation{Institut f\"ur Theoretische Physik, Universit\"at
  Hannover, D-30167 Hannover, Germany}
\affiliation{Grup de F\'isica Te\`orica, Universitat Aut\`onoma de Barcelona, E-08193 Bellaterra, Spain}
\author{Aditi Sen(De)}
\affiliation{Institut f\"ur Theoretische Physik, Universit\"at
  Hannover, D-30167 Hannover, Germany}
\affiliation{ICFO-Institut de Ci\`encies Fot\`oniques, Jordi Girona
  29, E-08034 Barcelona, Spain}
\author{ Ujjwal Sen}
\affiliation{Institut f\"ur Theoretische Physik, Universit\"at
  Hannover, D-30167 Hannover, Germany}
\affiliation{ICFO-Institut de Ci\`encies Fot\`oniques, Jordi Girona
  29, E-08034 Barcelona, Spain}
\author{Maciej Lewenstein}
\altaffiliation[Also at:]{ Instituci\'o Catalana de Recerca i Estudis
  Avan{\c c}ats.}
\affiliation{Institut f\"ur Theoretische Physik, Universit\"at
  Hannover, D-30167 Hannover, Germany}
\affiliation{ICFO-Institut de Ci\`encies Fot\`oniques, Jordi Girona
  29, E-08034 Barcelona, Spain}

\begin{abstract}

The logical structure of Quantum Mechanics (QM) and its relation 
to other fundamental principles of Nature has been for decades 
a subject of intensive research. In particular, the question 
whether the dynamical axiom of QM can be derived from other principles 
has been often considered. In this contribution, we 
show that unitary evolutions arise as a consequences of demanding 
preservation of entropy in the evolution of a single pure quantum
system, and preservation of entanglement in the evolution of
composite quantum systems.\footnote{\it { We would also like to dedicate this
work to the memory of Asher Peres, whose contributions and sharp comments
guided the first steps of the present article.}}

\end{abstract}

\maketitle

The standard axiomatic formulation of Quantum Mechanics (QM) 
relies on a set of postulates describing the state of a physical system, 
its time evolution, and the information that one can gather about the system 
by performing measurements. A specific formulation for joint
quantum systems - beyond the appropriate extension of the postulates -
is not regarded in such axiomatic approximation. However, one of   
the most genuine and counterintuitive properties 
of quantum mechanics - entanglement - appears only in joint quantum systems.
This quantum ``inseparability'' had for many decades questioned 
the logical structure of QM and raised a fundamental 
philosophical debate \cite{somebody}.
The most famous apparent ``paradox'' 
aroused  by the nonlocal character of composite quantum systems,
is exposed in the celebrated article of Einstein, Podolsky and Rosen~\cite{EPR} (the 
so-called EPR paradox), in which the authors claimed that in any plausible 
physical theory there exist ``elements of physical reality'' 
and questioned the completeness of QM
by analyzing entangled states. Afterwards  
Bell proved in his seminal contributions~\cite{Bell}
that if one assumes the validity of ``Einstein locality'', 
and an underlying hidden variable theory, there is an 
upper limit to the correlation between distant events. 
The claim, in itself, had nothing to do with QM, 
being a general statement about all physical theories
that have an underlying hidden variable model and respects Einstein locality. 
However, QM predicts the existence of certain entangled states 
whose correlations (as predicted by QM) violate 
the upper limit set by Bell (Bell inequalities). 
Violation of the Bell inequalities showed that
the conjunction of the 
principle of locality (Einstein locality) 
and the principle of realism (existence of a hidden
variable model) is incompatible with QM. After the discovery of the Bell 
inequalities, experiments were carefully performed to check the predictions  
of QM regarding the correlations in entangled states. 
However, the results of all the experiments performed till date 
to check for violation of Bell inequalities can be described 
by local hidden variable models, by taking advantage of the inefficiencies 
of experimental apparatuses. These effects have come to be known as 
``loopholes". E. Santos was one of the first to point this out \cite{loopholes} (see also \cite{others}, and references therein). 
Curiously enough, in the last decade, 
this puzzling character of entangled states has 
become the seed of an emerging technological revolution, whose consequences 
are hard to overestimate.\\ 
The aim of the present contribution is to establish a link between 
the axioms of QM and the consequences of demanding 
(i) {\bf{preservation of entropy}} in the evolution of a single pure quantum
system and (ii) {\bf{preservation of entanglement}} in the evolution of
composite quantum systems. We shall demonstrate that any of these requirements (either (i) or (ii)), 
together with other premises, 
retrieves the dynamical postulate of QM, 
stating that the evolution of a quantum 
system follows the Schr{\"o}dinger equation, i.e. the time evolution must 
be unitary ~\cite{CT}. In both cases, the converse is obviously true. 
The paper is organized as 
follows. In Sec. \ref{single-ek}, we briefly introduce
the axiomatic formalism of QM and our starting premises
to derive the dynamical postulate of QM from entropic assumptions and
probabilistic linearity. We study maps (transformations) 
that preserve disorder (entropy) and are linear 
on mixtures of density matrices. 
Sec. \ref{compo-dui} deals with preservation of 
entanglement in composite (bipartite) pure quantum systems. There, 
we investigate properties of linear maps which preserve 
entanglement either qualitatively or quantitatively. 
We demonstrate that every linear map
that preserves entanglement, has to be local, or local after a swap operation. 
We also show that if the preservation of entanglement is 
quantitative, then the map  is a local \emph{unitary}, 
or a multiple of a local unitary, after a swap operation.

\section{Single Quantum Systems} 
\label{single-ek}

\subsection{The Axioms of QM revisited }
\label{ek-dui}

The well established standard axiomatic formulation of QM
is based on four different classes of postulates (see e.g. \cite{CT}) 
dealing with description, measurements and evolution of a physical system. 
The first postulate concerns the ``static'' description 
of the system, asserting that to any physical system there is an 
associated Hilbert space whose dimension
corresponds to the number of ``degrees of freedom'' of the physical system. 
If complete information about the physical system is available, 
the state of the system is represented by an element 
of this Hilbert space (pure state). Otherwise, the state is represented by a 
density matrix, i.e. a self-adjoint, positive semidefinite, 
normalized  matrix acting on the Hilbert space. A second class of axioms of 
QM deals with the  measurements and their outcomes. There is a  
third type of axiom dealing with indistinguishably of identical particles
stating that bosons are represented by symmetric states in the corresponding 
Hilbert space while fermions by antisymmetric ones.  
These second and third type of axioms will not play any role 
in our considerations below. 
Finally, there exist a fourth type of axiom concerning the dynamics 
of a physical system. It states that 
the evolution (transformation from one state to another) 
of the system must be unitary. Thus, 
if a system evolves from a state 
\(\left| \phi \right\rangle\) to a state 
\(\left| \psi \right\rangle\), with \(\left| \phi \right\rangle\), \(\left| \psi \right\rangle\) 
\(\in {\cal H}\), then 
it must occur that 
 \(\left| \psi \right\rangle = U \left| \phi \right\rangle\),
where \(U\) is a unitary operator, i.e. \(U^\dagger U =I\) (\(I\) being  
the identity operator on the Hilbert space \({\cal H}\)).
If a physical system evolves from a 
density matrix \(\rho\) to a density matrix \(\rho{'}\),
then it should happen that \(\rho{'} = U \rho U^\dagger\).

The logical structure of QM, and its relation to other fundamental 
principles of Nature has been for decades a subject 
of intensive research 
(see e.g. \cite{EPR,Gleason,Bell,KS,cloning,GHZ123,Partovi,Hardy123,seijey1,
deleting,seijey56,Hardy,Dietz,Fuchs}). In particular, the question 
whether the dynamical axiom can be derived from 
other assumptions has been addressed in e.g. \cite{nonlinear}. 
In this paper, we give  an alternative derivation to 
show that this dynamical axiom can be derived from 
requiring that the evolution  
preserves disorder (as quantified by von Neumann entropy), in the spirit
of the second law of thermodynamics. Furthermore,
we will also consider a very weak assumption of probabilistic linearity on the 
dynamics of the system. 
More precisely, we consider maps $L$ fulfilling that:
\begin{itemize}

\item[(i)] the disorder of a physical system \(\rho\) 
is preserved in the evolution of the system transforming 
\(\rho\) to \(L(\rho)\).   The disorder of a physical system 
in a state \(\rho\) is quantified by its von Neumann entropy 
\(S(\rho)\), given by 
\begin{equation}
\label{pancha}
S(\rho) = - \Tr \rho \log_2 \rho.
\end{equation}

\item[(ii)] Probabilistic linearity: If two physical systems, described by states \(\rho_1\) and \(\rho_2\),
evolve to \(L(\rho_1)\)
  and \(L(\rho_2)\), then a physical system, described by a probabilistic mixture
\(p\rho_1 + (1-p)\rho_2\) (\(0 \leq p \leq 1\))
of \(\rho_1\) and \(\rho_2\), evolves to  
\(pL(\rho_1) + (1-p)L(\rho_2)\).

\end{itemize}
These are the only assumptions that we make on 
the dynamics of a physical system. We adopt here 
the ``static'' description of the physical system from Quantum Mechanics. 
But, by itself, this static description does not have any consequences 
on the dynamics of the system. 
Then we show that the evolution of the physical system can be 
either unitary or anti-unitary \cite{ulto}.
The converse of this statement is of 
course known: Unitary and anti-unitary evolutions  
preserve von Neumann entropy and can be defined 
to be probabilistically linear. As will be discussed later, 
anti-unitary operators cannot describe 
continuous transformations or better to say they cannot be
continously deformed to unitary operations.
Let us remark here that this result was previously 
obtained by A. Peres \cite{nonlinear}, showing that there is a close
relation between dynamical evolutions that violate some fundamental
axiom of QM, like unitarity, and those which are forbidden by the
second law of thermodynamics. We provide here an alternative 
derivation of this result.

\subsubsection{The assumption of preservation of disorder}
Let us first briefly discuss the question of quantification of disorder. 
Under some quite general assumptions, the amount of disorder 
of a probability distribution \(\{p_i\}\) can be shown to be the Shannon entropy
\(H(\{p_i\}) = -\sum_i p_i \log_2 p_i\). This quantifies, on average,  the amount of information 
an observer gathers (being equal to the disorder that was in the system)
when she/he gets to know the result of a random variable which is distributed according to
 \(\{p_i\}\). (See Ref. \cite{Welsh} for details.)
The above observation, along with the fact that 
 the eigenvalues of a density matrix \(\rho\)
are positive and sum up to unity  (i.e. they can be interpreted as probabilities), 
led von Neumann to define disorder in quantum mechanical systems as 
\(S(\rho)\), given by Eq. (\ref{pancha}),
for a physical system in a state \(\rho\).
Notice that
density matrices can 
 be interpreted as statistical mixtures of projectors onto elements of a Hilbert
space \({\cal H}\).
And \(S(\rho) = - \sum_{i=1}^n p_i \log_2 p_i\), 
where the $p_i$ are the eigenvalues of $\rho$, so that the 
von Neumann entropy of \(\rho\) is the Shannon entropy of the eigenvalues of  
\(\rho\). This is the only reason that we give for our choice of von Neumann entropy as 
the measure of disorder. For 
an axiomatic approach to derive the von Neumann entropy, as well as other reasons and advantages of this 
choice, see Ref. \cite{Wehrl}.
In passing, notice that the introduction of the von Neumann entropy does not require that the evolution is 
unitary, 
or anti-unitary \cite{vN1}. 
Note also that the assumption of preservation of 
disorder is not in contradiction with results of 
statistical mechanics. 
This point is further discussed in  Appendix \ref{sec:appA}.

\subsubsection{The assumption of probabilistic linearity}

We now move on to  discuss the assumption in item (ii).
As noted before, 
density matrices can 
 be interpreted as statistical mixtures of projectors 
onto elements of a Hilbert space \({\cal H}\). 
The probabilistic, or statistical interpretation of density matrices imply that a statistical mixture 
of two density matrices is a legitimate density matrix, implying   
by itself a ``probabilistic linearity'' of the dynamics. 
This probabilistic linearity of density matrices really 
employs the probabilistic (statistical) description 
of states, and that each element of the ensemble (that is described by a density matrix) evolves 
independently. 
This notion of linearity is used  e.g.  in classical mechanics and classical electrodynamics, 
when we deal with probabilistic mixtures of states of a system. This is one of the reasons, 
why we say that item (ii) is a ``weak'' assumption. 
Note that the probabilistic linearity is qualitatively different than the  
linearity of the quantum formalism on superpositions 
of vectors on a Hilbert space, \emph{which we do not assume}.
Note also that in our case, the evolution characterized by the map $L$ 
transforms a density matrix \(\rho\) into a matrix  which, in general, is not 
normalized.

\subsection{Evolutions that preserve disorder and are probabilistically linear are unitary}

The only states that have zero entropy are the pure states (one
eigenvalue equals unity and the others vanish). Therefore (by item (i)) every
pure state has to be mapped again to a pure state.
Consider two 
arbitrary different pure states $|\phi_1\rangle\langle\phi_1|\) and \(|\phi_2\rangle\langle
\phi_2|$. (The bras and kets used in this paper are normalized to unity.)
Since $L$ inevitably transform them to pure states we can write
\begin{eqnarray}
\label{Hannover}
L(|\phi_1\rangle \langle \phi_1|)&=& d_1 |\psi_1\rangle \langle \psi_1|, \nonumber \\
L(|\phi_2\rangle \langle \phi_2|)&=& d_2 |\psi_2\rangle \langle \psi_2|.
\end{eqnarray}
Although 
$L$ does not have to be
norm-preserving, 
\(L\) must transform density matrices into matrices, which after 
normalization are density matrices. Therefore \(d_1\) and \(d_2\) must be  both positive.
Otherwise, a statistical mixture of the outputs 
in Eq. (\ref{Hannover}), will not be a positive semidefinite matrix.

Consider now the transformation that takes \(|\phi_1\rangle\) to \(|\psi_1\rangle\) 
and \(|\phi_2\rangle\) to \(|\psi_2\rangle\).
Unless $|\phi_1\rangle$ is orthogonal to $|\phi_2\rangle$, and, 
since the choice is completely arbitrary, there must be a component 
of \(|\phi_2\rangle\) that is parallel to \(|\phi_1\rangle\), 
and a component that is orthogonal to 
 \(|\phi_1\rangle\). Thus,
there is a $|\Phi\rangle\) orthogonal to  
\(|\phi_1\rangle$ such that 
\begin{equation}
|\phi_2\rangle=\lambda_1 |\phi_1\rangle + \lambda_2
 |\Phi\rangle.
\end{equation}
We can 
draw the phase of
$\lambda_2$ into $|\Phi\rangle$,
so that we have $\lambda_2\geq 0$. 
The same argument applies to the $|\psi_i\rangle$. 
In this way 
\begin{equation}
\label{Sagor}
|\psi_2\rangle=\mu_1|\psi_1\rangle + \mu_2 |\Psi\rangle,
\end{equation}
with $\mu_2 \geq 0$, and \(|\Psi\rangle\) orthogonal to \(|\psi_1\rangle\).
Note that $|\lambda_1|=\sqrt{1-\lambda_2^2}$ and 
$|\mu_1|=\sqrt{1- \mu_2^2}$. The $|\lambda_i|$ and $|\mu_i|$ are now uniquely
defined and depend on the projectors $|\phi_i\rangle \langle \phi_i|$
and $|\psi_i \rangle \langle \psi_i|$. Note that \(\lambda_2\) and \(\mu_2\) 
cannot vanish.

Consider now the state 
\begin{equation}
\label{startstate}
\rho = p|\phi_1\rangle \langle \phi_1| + (1-p) |\phi_2\rangle \langle \phi_2|,
\end{equation}
with \(0\leq p \leq 1\). 
The matrix corresponding to $\rho$ in the  orthonormal basis $\{|\phi_1\rangle, 
|\Phi\rangle\}$ is 
\[
\left( \begin{tabular}{cc} $p+(1-p)|\lambda_1|^2$ & $(1-p)\lambda_1 \lambda_2$ \\
$(1-p) \lambda_1^*\lambda_2$ & $(1-p) \lambda_2^2$
\end{tabular} \right),
\] 
where the \(^*\) denotes a complex conjugation.
This matrix has the eigenvalues
\begin{eqnarray*}
s_1 &=& \frac{1}{2} (1 - \sqrt{1 - 4p \lambda_2^2 + 4p^2\lambda_2^2}), \\
s_2 &=& \frac{1}{2} (1 + \sqrt{1 - 4p \lambda_2^2 + 4p^2\lambda_2^2}).
\end{eqnarray*}
Due to probabilistic linearity (item (ii)),  $L$  maps $\rho$
 onto 
\begin{equation}
\label{endstate}
L(\rho)=
p d_1  |\psi_1\rangle \langle \psi_1| + (1-p) d_2 |\psi_2\rangle \langle 
\psi_2|.
\end{equation}
which has the eigenvalues
\[
t_1 = \frac{1}{2}(\alpha-\beta), \quad t_2 = \frac{1}{2}(\alpha+\beta),
\]
where 
\begin{eqnarray*}
\alpha &=& p d_1 + d_2 - p d_2, \\
\beta &=& \sqrt{(pd_1+d_2-pd_2)^2+4pd_1d_2\mu_2^2(p-1)}.
\end{eqnarray*}
For $2 \times 2$ 
matrices with unit trace, 
their von Neumann entropies can be equal if and only if 
they have the same eigenvalues.
Since $L$ has to be entropy preserving, the ratio 
between the eigenvalues of \(\rho\) 
and \(L(\rho)\) must 
be the same, so that
\[\mbox{either} \quad \frac{s_1}{s_2}=\frac{t_1}{t_2},
\quad \mbox{or} \quad \frac{s_1}{s_2}=\frac{t_2}{t_1}.\]

A short calculation yields
\begin{equation}
\mu_2=\frac{|p(d_1-d_2)+d_2|\lambda_2}{\sqrt{d_1d_2}}.
\end{equation}
However, 
$\mu_2$ was introduced in Eq. (\ref{Sagor}) and hence cannot depend on the mixing 
parameter $p$, which was introduced later on.
Therefore 
we must have, for all \(p\),
\begin{equation}
\frac{\vardelta \mu_2}{\vardelta p}=0,
\end{equation}
which implies that 
\[d_1=d_2.\] 
Hence, the two arbitrary pure states \(|\phi_1\rangle\) and \(|\phi_2\rangle\) are
mapped onto 
pure states
with the same length. Furthermore (using \(d_1 = d_2\)) it 
follows
that $\mu_2=\lambda_2$. But
this 
implies
that 
\[|\mu_1|=|\lambda_1|.\]
In other words,
the modulus of the scalar product is preserved in the evolution:
\begin{equation}|\langle \psi_1|\psi_2\rangle|=|\langle \phi_1|\phi_2\rangle|.\end{equation} 
The conservation of the modulus of the scalar product is, however, a very strong condition,
since using the  Wigner's theorem \cite{Wigner} (see Refs. \cite{Emch,Gisin}
in this regard), we obtain now  that the transformation that 
induces
\begin{eqnarray}
\label{abar}
|\phi_1\rangle & \rightarrow & |\psi_1\rangle, \nonumber \\
|\phi_2\rangle & \rightarrow & |\psi_2\rangle,
\end{eqnarray}
is either linear and unitary or antilinear and anti-unitary. 
Here linearity means that if a transformation induces the transitions in Eq. (\ref{abar}), then 
the same transformation induces the transition
\(a|\phi_1\rangle
+ b|\phi_2\rangle \rightarrow
a |\psi_1\rangle
+ b |\psi_2\rangle\). And antilinearity means that the obtained vector is 
 \(a^* |\psi_1\rangle
+ b^* |\psi_2\rangle\).

Continuous transformations cannot be described by antilinear operators, as two 
consecutive antilinear operators act as a linear operator. We will disregard the option of 
anti-unitary operators for this reason. The evolution of our physical system 
is not given by Eq. (\ref{abar}), instead, it is given by 
Eq. (\ref{Hannover}). However, we 
have shown that \(d_1 = d_2\). Thus, the evolution on our physical system is 
unitary, up to a constant. 
As this constant is independent of the input state, it is 
irrelevant.  Therefore we have reached our goal. 
The evolutions that respect the second law of thermodynamics
and are probabilistically linear, are just the ones which are postulated in 
Quantum Mechanics: linear (on 
superpositions of vectors) and unitary.

\section{Composite Quantum Systems}
\label{compo-dui}

In this second part of the paper, we extend our study to bipartite 
(pure) quantum  systems. We restrict our study to maps that 
are linear (on superpositions of pure states) and preserve entanglement. 
We will show that demanding qualitative preservation of entanglement, 
i.e. mapping separable states onto separable states and
entangled states onto entangled states requires the map to be
local or the product of a local map and the swap operator. The
stricter requirement of preservation of some measure of entanglement
requires the local maps to be essentially unitaries. 

At this point, it is worth stressing the difference between the linearity postulates, assumed in this paper,
for single and composite quantum systems. In Sec. \ref{single-ek}, we assumed a probabilistic linearity on 
the evolution maps, which is a rule for \emph{mixtures} of density operators. In this section, on the other hand,
evolution maps are  
assumed to  be linear on \emph{superpositions} of pure states. Henceforth, the latter version is simply 
referred to as ``linear'' maps.

\subsection{Formalism}
\label{sec:formalism}

We restrict our discussion to bipartite systems, traditionally called
 Alice and Bob. The Hilbert space associated to the bipartite system
is denoted by $ {\cal H}_{AB}:= {\cal H}_A \otimes {\cal H}_B$,
where ${\mathcal H}_{A/B}$ are Hilbert spaces of dimension $n$, 
respectively $m$, over the field $\field$. We restrict ourselves to 
linear maps and allow the addition of local ancillas. The structure of the maps
considered here is the following: 
\begin{equation}
L : {\mathcal H}_{AB} \mapsto {\mathcal H}_{A} \otimes {\mathcal H}_{Aa}
\otimes {\mathcal H}_{B} \otimes {\mathcal H}_{Ba}.
\end{equation}
The above map $L$ could be decomposed into two different maps, the first one
consisting only in adding local ancillas to the initial state, while the second
one maps this state into the final one: 
$L:{\cal H} \hookrightarrow {\cal H} \otimes {\cal H}_{Aa} \otimes
{\cal H}_{Ba} \xrightarrow{L'} {\cal H} \otimes {\cal H}_{Aa} \otimes
{\cal H}_{Ba}$.  
The first step is obviously entanglement preserving since the ancillas are
in a product state. Thus for $L$ to be entanglement
preserving, it is enough to demand  $L'$ to be entanglement preserving on the
image of the inclusion. This means that $L'$ acting on any state of the
form $|\psi\rangle\otimes |A_a\rangle \otimes B_b\rangle$ is entanglement
preserving, but there might be more general states in ${\cal H}
\otimes {\cal H}_{Aa} \otimes {\cal H}_{Ba}$ on which $L'$ is not
entanglement preserving.

By dimensional arguments ($\dim (\range L) 
\leq \dim {\mathcal H}_{AB}=nm$) one can easily realise
that there exist states in ${\calhanc}$ which do not correspond to 
the image of $L$, i.e. there exist elements of ${\calhanc}$ which are
not reached by mapping the elements of ${\mathcal H}_{AB}$  with $L$.
We can, therefore, restrict our investigation to a $nm$-dimensional subspace 
${\mathcal R} \subset \calhanc$
into which all original states from ${\cal H}$ are mapped. 
Thus, mathematically we study 
linear, entanglement preserving maps of the form:
\begin{equation}
L:{\mathcal H}_A \otimes {\mathcal H}_B \mapsto {\mathcal R}.
\end{equation}
We will show later that in fact ${\mathcal R}$ is isomorphic to ${\mathcal H}_{AB}$.
In the most general case the map $L$ is not necessarily norm preserving.
To keep track of this fact, as in Sect. I  we use the ``bra"
and ``ket" notation only for normalized states, i.e. $\langle
\psi|\psi\rangle=1$. We shall explicitly write down any
non-unit-length pre-factors (i.e.\ $c |\psi\rangle$). A map
that decreases the norm of some states can be physically interpreted as a
device that prepares the final state with a non-unit probability.
In the remaining cases, the device prepares no system for the output.
Norm-increasing maps are harder to interpret but we allow
them for completeness.
Finally, in our notation, if two states are parallel, i.e. if for
a given $|\psi\rangle$ and $|\phi \rangle$, there 
exists a $c \in \field$ such that $|\psi\rangle =
c|\phi\rangle$, we use the notation
$|\psi\rangle \parallel |\phi \rangle$. Non parallel states will be denoted
by $|\psi\rangle \nparallel |\phi \rangle$.

\subsection{Entanglement preservation}
\label{sec:entangl-pres}

To characterize all maps that preserve entanglement, we have to first define
this property (of entanglement).  In the following, we distinguish between
\emph{qualitative entanglement preservation} and the stronger
\emph{quantitative entanglement preservation}. Qualitative
entanglement preservation requires that separable
states are mapped to separable states and entangled states are mapped
to entangled states. This type of entanglement preservation leads to the fact
that the map must be either local, or the product of a local map 
times the \emph{swap operator}.

For a map to be quantitatively entanglement preserving, we should demand that
there exists an entanglement measure $E$ \cite{entanmeasure,onek}, such that
for all initial states $|\psi\rangle$ it holds 
$E(|\psi\rangle)=E(L(|\psi\rangle) )$.  
It is known that there exists exactly one asymptotic measure of entanglement
for pure normalized bipartite states, which is the von Neumann entropy
of the reduced system of one of the two parties \cite{horo99}. In the
remainder of this paper, we will denote this quantity by
$E(|\psi\rangle)$. Using na\"ively the entropy function also on
non-normalized states does not lead to a sensible entanglement
measure.  The entropy function evaluated on $|c|^2
|a\rangle\langle a|$ (the reduced state of the pure product state
$c |ab\rangle$) gives $-|c|^2\log_2 |c|^2$, which is 0
only for $|c| \in \{0,1\}$.  There are, however, different ways
to extend $E(|\psi\rangle)$ towards non-normalized states and in this
paper we will deal with two possibilities: The first possibility is to
normalize the state, i.e., to use the \emph{renormalized measure}
$E_1(c|\psi\rangle)=E(|\psi\rangle)$.  This measure nullifies any
physical significance that we may try to ascribe to a possible
norm-change under the map.  Alternatively, we may keep the norm as a
multiplicative pre-factor, i.e., we use the \emph{probabilistic
  measure} $E_2(c|\psi\rangle)=|c|^2 E(|\psi\rangle)$. In
the ``failed map'' interpretation of norm-loss, this measure is the
average entanglement the device produces in the long run.

Of course, quantitative entanglement preservation  
implies qualitative entanglement preservation. But it is the 
quantitative preservation which allow us to demonstrate that 
such a map is, essentially, a local
unitary. Before going into details, we review 
for completeness, some well established concepts such as the 
Schmidt decomposition for bipartite pure states and the swapping operator.

Bipartite systems always admit a 
\emph{Schmidt-decomposition}, i.e. for every pure state $|\psi\rangle
\in {\cal H}_{AB}$, there exists orthonormal bases $\{|a_i\rangle\}$ of
${\cal H}_A$ and  $\{|b_i\rangle\}$ of ${\cal
  H}_B$, such that $|\psi\rangle= \sum_{i=1}^r \lambda_i |a_i\rangle
\otimes |b_i\rangle$, with $\lambda \geq 0$ \cite{Schmidt}. The
$\lambda_i$ are denoted as the {\em Schmidt coefficients} and the $r$
as the {\em Schmidt rank} of the state $|\psi\rangle$. 
Hence the Schmidt rank is the minimal number of product states needed to
decompose $|\psi\rangle$, and, by construction, is bounded by
$r\le \min(n,m)$.

Finally, we define here the {\em swap} operator as 
an special map that preserves entanglement. Swapping, in this paper, corresponds to 
exchange or relabeling the two subsystems and is mathematically a map
$S:{\mathcal H}_A \otimes {\mathcal H}_B \mapsto {\mathcal H}_{A'}
\otimes {\mathcal H}_{B'}$, where ${\mathcal H}_{A'}$ is isomorphic to
${\mathcal H}_B$ and ${\mathcal H}_{B'}$ is isomorphic to ${\mathcal
  H}_A$.  For a given orthonormal basis $\{|{\mathbf 1}_A\rangle,...,
|{\mathbf n}_A\rangle\}$ of ${\mathcal H}_A$ and $\{|{\mathbf
  1}_B\rangle,..., |{\mathbf m}_B\rangle\}$ of ${\mathcal H}_B$, the
isomorphisms immediately lead to orthonormal bases $\{|{\mathbf
  1}_{A'}\rangle,..., |{\mathbf m}_{A'}\rangle\}$ of ${\mathcal
  H}_{A'}$ and $\{|{\mathbf 1}_{B'}\rangle,..., |{\mathbf
  n}_{B'}\rangle\}$ of ${\mathcal H}_{B'}$. We define the swap
operator as
\begin{equation}
\hat{S}:|{\mathbf i}_A\rangle \otimes |{\mathbf j}_B\rangle \mapsto 
|{\mathbf j}_{A'}\rangle \otimes |{\mathbf i}_{B'}\rangle.
\end{equation}
Of course, a different isomorphism between ${\cal H}_{A/B}$ and 
${\cal H}_{A'/B'}$, that 
identifies the bases $\{|{\bf i}_{A/B}\rangle\}$ with 
different orthonormal bases $\{|{\bf \tilde{i}}_{A'/B'}\rangle\}$ leads to a
different swap operator. But since
all local orthonormal bases
are connected by local unitaries, all swap operators are also interconnected 
by local unitaries, that is, up to local unitaries, the swap operator is unique.
Note here that the swap operator considered in this paper is different from the usual swap operator 
considered in the literature, which also allows exchanging a subspace of the Hilbert space 
\({\mathcal H}_A\) with that of \({\mathcal H}_B\). Such a general swap operator can of course 
increase (or  decrease) the entanglement between \(A\) and \(B\): Simply consider exchanging 
\(A''\) with \(B''\) (and back) for the state \(|\Psi^{-}\rangle_{A'A''} |0\rangle_{B'} |0\rangle_{B''}\) in 
the \(A'A'': B'B''\) bipartite cut, where \(|\Psi^{-}\rangle =  \frac{1}{\sqrt{2}}(|01\rangle - |10\rangle)\).

\subsection{Qualitative entanglement preservation}
\label{sec:weak_premiss}
Qualitative entanglement preservation is equivalent
to the demand that the set of states with Schmidt rank $r=1$
(product states) and the set of states with Schmidt rank $r>1$
(entangled states) are both invariant under application of the map $L$.
\begin{fact}
  \label{fact:rank-entangl-pres} A qualitatively entanglement preserving linear map has to be of full
  rank.
\end{fact}

Proof:  Suppose that the kernel of a linear 
qualitatively entanglement preserving map \(L\) is not empty. 
Assume that there exist a product vector $\ket{a_1,b_1}$  which 
belongs to the kernel of \(L\), i.e. $L\ket{a_1,b_1}=0$. 
Let us choose a second product vector $\ket{a_2,b_2}$,
for which $\ket{a_2}$, $\ket{b_2}$ are not parallel 
to $\ket{a_1}$, $\ket{b_1}$ respectively.
Any (nontrivial) combination of these two
product states \(|\psi_2\rangle\), is an 
entangled state of Schmidt rank 2. But unless $\ket{a_2,b_2}$ 
is also in the kernel, $L$ cannot be
entanglement preserving. Thus, if there exist one product state 
in the kernel, then one can find a basis of product states 
all of them belonging to the kernel of $L$, and hence $L$ is the zero map. 
It is enough to show now that there exit no entangled state of
Schmidt rank 2 in the kernel of $L$. 
Assume that there exist one entangled state of rank 2  
$\ket{\phi^+_2}=\lambda_1\ket{e_1,f_1}+\lambda_2\ket{e_2,f_2}$,  written in 
the Schmidt decomposition with $\lambda_i>0$ in the kernel of $L$, i.e. 
$L(\ket{\phi^+_2})=0$. We construct 
$\ket{\phi^-_2}=\lambda_1\ket{e_1,f_1}-\lambda_2\ket{e_2,f_2}$ which is also
of rank 2. However, $(1/2)(\ket{\phi^+_2}+\ket{\phi^-_2})=\lambda_1\ket{e_1,f_1}$ and therefore
$L(\ket{\phi^-_2})$ must be of rank 1, which cannot be since $L$ is 
entanglement preserving map. Therefore, there 
exist not entangled state of rank $r=2$ in the Kernel of $L$. 
It is now obvious to see that if the kernel of $L$ does not contain 
any product vector and any Schmidt rank 2 vector, 
then it cannot contain any vector of 
Schmidt rank $r>2$. Therefore, we have shown that if $L$ is a non trivial 
entanglement preserving map, its kernel is empty and $L$ is of full rank.

We observe that $L$ maps all product
states to multiples of product states. Therefore, $L$ cannot
increase the Schmidt rank of a pure state: If $|\phi\rangle=
\sum_{i=1}^r \lambda_i |a_i b_i\rangle$, then $L|\phi\rangle= \sum_{i=1}^r
\lambda_i L|a_i b_i\rangle= \sum_{i=1}^r\lambda_i c_i |a'_i
b'_i\rangle$ is a superposition of at most $r$ product terms, and the
Schmidt rank of $L|\phi\rangle$ cannot be larger than $r$.
Furthermore, we know from Fact~\ref{fact:rank-entangl-pres} that $L$
is invertible,
that is, for each $L$ that is entanglement preserving, 
there exists an $L^{-1}$ with the property that
$L^{-1}L|\phi\rangle= |\phi\rangle$.
Thus $L^{-1}$ is also a linear qualitatively entanglement preserving map.
Thus neither $L$ or $L^{-1}$ can increase the Schmidt rank. So, starting from
the weak demand of qualitative entanglement preservation we have proved
\begin{fact}
The Schmidt rank is invariant 
under the application of any linear, qualitatively entanglement preserving map $L$.   
\end{fact}


We are now ready to state our main result regarding qualitative
entanglement preservation.
\begin{result}
  \label{eq:ent_pre-loc}
  Every linear map that is qualitatively entanglement preserving has
  to be either local or a local operator times the swap operator.    
\end{result}
The first part of our proof of this result is slightly technical, and
proceeds by looking at some specific states and their image under an
entanglement preserving map. To do this, we begin by choosing an orthonormal basis 
$\{|{\mathbf
  1}_{A}\rangle,...,|{\mathbf n}_A\rangle\}$ of
Alice's Hilbert space ${\mathcal H}_A$, and an  orthonormal basis $\{|{\mathbf 1}_B \rangle,...,|{\mathbf
  m}_B\rangle\}$
of Bob's Hilbert space ${\mathcal H}_B$. Due to the preservation of separability, we know that
$L$ acts on an element of the product basis $|{\bf i}_A,{\bf j}_B
\rangle$ as:
\begin{equation}
\label{l-def-eq}
|{\mathbf i}_A,{\mathbf j}_B \rangle \xrightarrow{L} c_{i,j}
|d_{i,j}\rangle \otimes |e_{i,j} \rangle,
\end{equation}
where the length $c_{i,j}$ and both tensor factors can a priori depend
on both input factors $|{\bf i}_A\rangle$ and $|{\bf j}_B\rangle$.
Since $L$ is linear, this evaluation on a basis fixes $L$ completely.
Now we start by applying $L$ to a state of Schmidt rank 2. Let $ i \neq k$
and $ j \neq l$ and we look at the transformation
\begin{equation}
  |{\mathbf i}_A{\mathbf j}_B\rangle + |{\mathbf l}_A{\mathbf
      k}_B\rangle 
  \xrightarrow{L}
  c_{i,j} | d_{i,j}\rangle \otimes |e_{i,j}\rangle +
      c_{k,l}|d_{k,l}\rangle \otimes |e_{k,l}\rangle.
\end{equation}
As the right hand is not a product state, we must have
$|d_{i,j}\rangle \nparallel |d_{k,l}\rangle$ and $|e_{i,j}\rangle \nparallel
|e_{k,l}\rangle$, i.e., when both indices differ, the image vectors
cannot be parallel. On the other hand, consider the following application of the map 
to a product vector:
\begin{align}
  |{\mathbf 1}_A{\mathbf 1}_B\rangle +|{\mathbf 1}_A{\mathbf 2}_B\rangle= 
  |{\mathbf 1}_A\rangle \otimes [ |{\mathbf 1}_B\rangle + |{\mathbf 2}_B
  \rangle]  \nonumber \\
\xrightarrow{L} c_{1,1}
  |d_{1,1}\rangle \otimes |e_{1,1}\rangle +
  c_{1,2} |d_{1,2}\rangle \otimes |e_{1,2}\rangle
\end{align}
It is clear that either
(i) $|d_{1,2} \rangle \parallel |d_{1,1}\rangle$, or
(ii) $|e_{1,2} \rangle \parallel |e_{1,1}\rangle$.

It turns out that in case (i),
$|d_{i,j}\rangle
\parallel |d_{i,i}\rangle$ and $|e_{i,j}\rangle \parallel
|e_{j,j}\rangle$, while in case (ii), it is the other way around: the
$|d\rangle$ image vectors only depend on the second index, while the
$|e\rangle$ image vectors only depend on the first index. 

To prove this, assume that we know that $|d_{i,j} \rangle \parallel
|d_{i,l}\rangle$ for $j\neq l$ and we look at
\begin{align}
  |{\mathbf i}_A{\mathbf j}_B\rangle +|{\mathbf k}_A{\mathbf j}_B\rangle= 
  \left[ |{\mathbf i}_A\rangle + |{\mathbf k}_A\rangle\right] 
  \otimes
  |{\mathbf j}_B\rangle  \nonumber \\
  \xrightarrow{L} c_{i,j}
  |d_{i,j}\rangle \otimes |e_{i,j}\rangle +
  c_{k,j} |d_{k,j}\rangle \otimes |e_{k,j}\rangle,
\end{align}
for some $k \neq i$. On the right hand side, the $d$--factors cannot be
parallel, since $|d_{i,j}\rangle \parallel |d_{i,l}\rangle$ and
$|d_{i,l}\rangle \nparallel |d_{k,j}\rangle$ as we pointed out before since they differ in both
indices. Therefore we find that
\begin{equation}
  \label{eq:parcon_d}
\exists l \neq j:
|d_{i,j} \rangle \parallel
|d_{i,l}\rangle 
\quad
\Rightarrow
\quad
|e_{i,j}\rangle \parallel |e_{k,j}\rangle
\quad \forall k,
\end{equation}
and in an analogous way we can prove that
\begin{equation}
\label{eq:parcon_e}
\exists k \neq i:
|e_{i,j} \rangle \parallel
|e_{k,j}\rangle 
\quad
\Rightarrow
\quad
|d_{i,j}\rangle \parallel |d_{i,l}\rangle
\quad \forall l
.
\end{equation}
Using $|d_{1,2} \rangle \parallel |d_{1,1}\rangle$ as the starting point,
it is now easy to show the desired result for case (i):
\begin{align}
  \label{eq:from_d12_to_rest}
  |d_{1,2} \rangle \parallel |d_{1,1}\rangle
  \quad
  \Rightarrow 
  \quad
  |e_{i,1} \rangle \parallel |e_{1,1}\rangle \quad \forall i
\nonumber \\  
\Rightarrow
  \quad
  |d_{i,j}\rangle \parallel |d_{i,i}\rangle \quad \forall i,j
  \quad
  \Rightarrow
  \quad
  |e_{i,j} \rangle \parallel |e_{j,j}\rangle \quad \forall i,j
  ,
\end{align}
which means that
\begin{equation}
\label{eq:case1}
|{\mathbf  i}_A{\mathbf  j}_B\rangle
\xrightarrow{L}
c_{i,j}|d_{i,j}\rangle \otimes 
|e_{i,j}\rangle=\tilde c_{i,j}
|d_{i,i}\rangle |e_{j,j}\rangle
,
\end{equation}
where $\tilde c_{i,j}$ may contain extra phase factors as compared to
$c_{i,j}$. Therefore we have obtained a decomposition of $L$ in the form
\begin{equation}
L= L_A \otimes L_B P,
\end{equation}
where $L_{A/B} : {\mathcal H}_{A/B} \mapsto {\mathcal H}_{A/B}$,
with $L_A|{\mathbf i}_A\rangle=|d_{i,i}\rangle$ and $L_B|{\mathbf
  j}_B\rangle= |e_{j,j}\rangle$ and $P$ is a global phase/length with
$P:{\mathcal H}_{AB} \mapsto {\mathcal H}_{AB}$ with $P|{\mathbf
  i}_A{\mathbf j}_B \rangle \mapsto \tilde c_{i,j} |{\mathbf
  i}_A{\mathbf j}_B\rangle$. Obviously, the map $P$ is diagonal in the
chosen product basis, but since  $L_A \otimes L_B$ is local, it is also
clear that $P$ must map \emph{all} product vectors to product
vectors in order to fulfill the preservation of separability.
For case (ii), Eqs.~(\ref{eq:parcon_d}) and (\ref{eq:parcon_e}) must be
replaced by their equivalents with $d$ and $e$ swapped. Instead of
Eq.~(\ref{eq:case1}), we then obtain
\begin{equation}
\label{eq:case2}
|{\mathbf  i}_A{\mathbf  j}_B\rangle
\xrightarrow{L}
c_{i,j}|d_{i,j}\rangle \otimes 
|e_{i,j}\rangle=\tilde c_{i,j}
|d_{j,j}\rangle |e_{i,i}\rangle 
.
\end{equation}
Since the set $\{|{\mathbf  i}_A{\mathbf  j}_B\rangle\}$ spans
${\mathcal H}_{AB}$,  and $L$  
is of full rank, the image of these vectors, i.e. the 
set $\{|d_{j,j} \rangle |e_{i,j}\rangle\}$, has also to span the 
whole space ${\mathcal R}$, which is $nm$ dimensional. So both sets
$\{|d_{j,j}\rangle\}$ and $\{|e_{i,i}\rangle\}$ have to be linearly
independent, and span a $m$ dimensional Hilbert space ${\mathcal
  H}_{A'}$, respectively a $n$ dimensional Hilbert space ${\mathcal
  H}_{B'}$, so that 
\begin{equation}
{\mathcal R}= {\mathcal H}_{A'} \otimes 
{\mathcal H}_{B'}.
\end{equation}
In this case, the combined map $S \circ L$, where $S$ is the swap operator,
maps from the space ${\mathcal H}_{AB}$
into itself (or more specifically in a space isomorphic to 
${\mathcal H}_{AB}$), and can be decomposed as 
\begin{equation}
S \circ L = L_A \otimes L_B P
,
\end{equation}
 where 
$L_A |{\bf i_A}\rangle = |e_{i,i}\rangle$, $L_B|{\bf j_B}\rangle =
|d_{j,j}\rangle$ and $P |{\bf i_A j_B}\rangle= \tilde c_{i,j}
|{\bf i_A j_B}\rangle$. As in the first case, $P$ maps all
product vectors to product vectors.

To complete the proof of
Result~\ref{eq:ent_pre-loc} we now only need
\begin{fact}
\label{fact3}
Every linear map that takes all product states to product
  states, is of full rank, and is diagonal in some product basis, has to be local.
\end{fact}
To prove this, we proceed as follows.
Let $P$ be of diagonal form in the product basis $|a_i b_j\rangle$, that is
\begin{equation}
P|a_ib_j\rangle=\lambda_{ij} |a_i b_j\rangle
.
\end{equation}
We know that $P$ maps the product state
$(\sum_i \alpha_i |a_i\rangle)\otimes(\sum_j \beta_j |b_j\rangle)$ 
(where we take the $\alpha_i \not = 0 \not = \beta_j$ as arbitrary but fixed)
to a product state, which can in general be written as 
 $\sum_{i,j} x_i y_j |a_i b_j\rangle$. 
This can only hold if for all $i$ and $j$
\begin{equation}
x_i y_j= \lambda_{ij} \alpha_i \beta_j.
\end{equation}
Since $P$ is of full rank, all
$\lambda_{ij} \not = 0$, and thus, it also has to hold that 
$x_i y_j \not = 0$ and we can write
\begin{equation}
\label{yfrac}
\frac{y_j}{y_{k}}
= 
\frac{\lambda_{ij}}{\lambda_{ik}} \frac{\beta_j}{\beta_{k}}.
\end{equation}
The left hand side of eq (\ref{yfrac}) does not depend on $i$, so 
the fraction 
$\chi_{jk}:= \frac{\lambda_{ij}}{\lambda_{ik}}$
does neither. This gives us a separation of the form
\begin{equation}
\lambda_{ij}=\lambda_{i0} \chi_{j0} =: \mu_i \nu_j.
\end{equation}
Therefore $P=P_A \otimes P_B$ as stated above, where $P_A
|a_i\rangle=\mu_i |a_i\rangle$ and $P_B |b_j \rangle = \nu_j
|b_j\rangle$.

\subsection{Quantitative entanglement preservation}
\label{sec:strong_premiss}
In the previous section, we only required a qualitative preservation
of entanglement. In this section, we will discuss the quantitative
preservation of entanglement. As stated in Sec.~\ref{sec:entangl-pres},
a quantitative preservation of entanglement depends on the extension of the
von Neumann entropy of the reduced system that is used.  However, for
both of the two previously defined entanglement measures, $E_1$ and
$E_2$, we have that
\begin{result}
  \label{result:quan_loc_uni}
  Every local and linear map $L$ that is quantitatively entanglement
  preserving, has to be a multiple of a local unitary.
\end{result}
We have proved in the previous section that a qualitatively entanglement 
preserving map is a local map or a local map after the application of the swap 
operator. Every quantitatively entanglement preserving map is also a 
qualitatively entanglement preserving one, so we know that 
$L=L_A \otimes
L_B$. Due to the singular value decomposition of $L_A$ and $L_B$ we
have
\begin{align}
\label{sing1}
L_A= V_A D_A U_A, & &
L_B= V_B D_B U_B,
\end{align}
where $U_A, U_B, V_A, V_B$ are unitary matrices, $D_A$ is a diagonal
matrix in the basis $\{ |{\bf 1}_A\rangle,...,|{\bf n}_A\rangle\}$ 
with the
diagonal values $ \lambda_1 \geq \lambda_2 \geq... \geq \lambda_n >0$,
and $D_B$ is a diagonal matrix in the basis
$\{{\bf 1}_B\rangle,...,|{\bf m}_B\rangle\}$ with the diagonal 
values $\mu_1 \geq
\mu_2 \geq ... \geq \mu_m >0$.  We define four further sets of
orthonormal bases by
\begin{subequations}
\begin{align}
|a_i\rangle&:=U_A^{-1}|i_A\rangle,  &|b_i\rangle:=U_B^{-1}|i_B\rangle, \\
|e_i\rangle&:=V_A|i_A\rangle,  &|f_i\rangle:=V_B|i_B\rangle.
\end{align}
\end{subequations}
For the respective proofs we will deal with a special set of states defined as
\begin{equation}
\label{vorl}
|\psi(c) \rangle= c |a_1 b_1\rangle + \sqrt{1-|c|^2}
|a_n b_m\rangle)
,
\end{equation} 
where $c\in[0,1]$. These states transform under $L$ into
\begin{equation}
\label{nonmax}
L | \psi(c) \rangle = c\lambda_1 \mu_1 |e_1
f_1\rangle + \lambda_n \mu_m\sqrt{1-|c|^2} |e_n f_n\rangle.
\end{equation}
Note that Eq. (\ref{nonmax}) expresses $L |\psi\rangle$ already in its
Schmidt decomposition.

The way to proceed now depends on the entanglement measure we choose.

\subsubsection{The renormalized measure of entanglement $E_1$}

In this subsection we will use the entanglement measure
$E_1(\kappa|\psi\rangle):= E(\psi\rangle)$.  Let us look at the state
$|\psi(c_1)\rangle$ of (\ref{vorl}) with $c_1=1/\sqrt{2}$. This is obviously a
maximally entangled state 
with Schmidt rank $2$:
$E_1(|\psi(c_1)\rangle)=E(|\psi(c_1)\rangle)=1$. 
The state~(\ref{nonmax}) has this entanglement value only if
$\lambda_1\mu_1 = \lambda_n \mu_m$, which fixes $\lambda_n=\lambda_1$
and $\mu_m=\mu_1$. (We remember that the $\mu$'s and $\nu$'s are
ordered.) So $D_A=\lambda_1 \eins_A$ and $D_B= \mu_1 \eins_B$, which
then leaves us with
\begin{align}
\label{loc-unit}
L_A= \lambda_1 V_A U_A, & &
L_B= \mu_1 V_B U_B, 
\end{align}
which are multiples of products of unitary matrices. 
This concludes the proof of
Result~~\ref{result:quan_loc_uni} for the renormalized entanglement
measure.

\subsubsection{The probabilistic measure of entanglement $E_2$}

We will now study the entanglement measure $E_2(c
|\psi\rangle)=|c|^2E(|\psi\rangle)$. Let us start with the physically most
reasonable case, namely $\lambda_1\mu_1 \leq 1$. Assume now that we
actually have $\lambda_n\mu_m <1$. Then the entanglement of the 
maximally entangled state
$|\psi (\frac{1}{\sqrt{2}})\rangle= 
\frac{1}{\sqrt{2}} (|a_1b_1\rangle +|a_nb_m\rangle)$
cannot be preserved, since this state would be mapped to the
multiple of a state that is at most as entangled as the input state,
but the length (or the probability of getting this state) would be
$\sqrt{\frac{\lambda_1\mu_1^2}{2}+\frac{\lambda_n \mu_m^2}{2}}<1$, 
leading to an even smaller actual value of the
probabilistic measure of entanglement. Therefore we would have in
this case
$\lambda_1=\lambda_n=1=\mu_1=\mu_m$, and $L$ is unitary.

Since $L$ is of full rank, it is invertible and $L^{-1}$ is also
entanglement preserving, so that similar arguments can be used to deal
with the case $\lambda_n \mu_m \geq 1$. The only remaining
case is then $\lambda_1\mu_1 \geq 1 \geq \lambda_n\mu_m$, and we deal with
that in Appendix~\ref{sec:appB}, since it is rather 
technical, and moreover this case involves the rather
unreasonable argument of the increase in norm for some states.

\section{Conclusion}
\label{sec:conclusion}

Summarizing, we have investigated the properties of maps that preserve
either entropy in single quantum systems or entanglement in composite
quantum systems. The first part of the paper demonstrates that the  
evolutions  that respect the second law of thermodynamics
and are probabilistically linear, are just the ones which 
are postulated in 
Quantum Mechanics: linear (on 
superpositions of vectors) and unitary. 
In the second part of the paper we have shown, 
that if from a linear  map \(L\) (linear on superpositions of vectors), 
we additionally demand
preservation of entanglement 
in a very reasonable form (precisely that product states are mapped to product
states and entangled states to entangled ones), 
then one immediately obtains preservation of the Schmidt rank. 
From this fact we have concluded
that every reasonable entanglement preserving map has to be either 
local or local after application of a swap operation.  
In a further step, we have studied the consequences that two 
special extensions of the unique asymptotic entanglement measure 
for pure bipartite states, induce on the two local parts of $L$.  
We were able to show that for the probabilistic measure of entanglement $E_2$,
the linear map $L$ is in fact a local \emph{unitary}, 
or a local unitary after the 
application of the swap. And in the
case of the renormalized measure of entanglement $E_1$,  
the map is either a multiple of a
local unitary, or a multiple of a local unitary after the application
of the swap. But, in this case (of \(E_1\)), we can drop the factor, since we
renormalize the outcome, for the evaluation of the entanglement.
Furthermore, we have shown along the way, that the addition of the local
ancillas does not increase the set of allowed maps since the space
mapped into has to remain isomorphic to the composite Hilbert space ${\mathcal H}_{AB}$.

It is tempting to think of our results 
in the following way: entropy is, of course, preserved in unitary dynamics. 
However, for systems whose dynamics is not assumed to
be quantum, entropy preservation can be seen as the second law of
thermodynamics (for reversible processes). 
On the other hand, for composite systems,
it is clear that entanglement is preserved by local unitary dynamics 
and (complete) exchange of systems. 
However, for composite systems on which we do not put
any dynamical assumption, 
preservation of entanglement has been proposed as an equivalent
second law of thermodynamics for composite systems (see e.g.~\cite{thermo}). 
The premises we use could, therefore, be regarded as 
second laws for single and composite
systems plus some forms of linearity on the evolution. 
And under these assumptions we have shown that the evolution must be
unitary.

\begin{acknowledgments}
We would like to thank William K. Wootters 
for illuminating correspondence,
Dagmar Bru{\ss} and Klaus Dietz for discussions, 
and the participants of the A2 Consortium.
We acknowledge support from the 
DFG (SFB 407, SPP 1078 ``Quantum Information Processing'' and the 
European Graduate College ``Interference and Quantum Applications'', 432POL), 
the Alexander von Humboldt Foundation, the EC program QUPRODIS, and the ESF Program QUDEDIS.
\end{acknowledgments}

\appendix

\section{}
\label{sec:appA}

In this Appendix, we argue that the 
assumption of preservation of disorder in reversible 
processes (item(i) of \ref{ek-dui})
can coexist 
with statistical mechanics, in particular with the Boltzmann \(H\) theorem.

Suppose that there are \(N\) classical particles of a gas in a box, such that 
the probability density at position \(\vec{r}\), momentum \(\vec{p}\), and time \(t\), is given by the distribution
function 
\(f(\vec{r}, \vec{p}, t)\). Then if the particles follow classical mechanical equations, the quantity 
\[
H(t) = - \int d\vec{r} d\vec{p} f(\vec{r}, \vec{p}, t) \log_2 f(\vec{r}, \vec{p}, t)
\]
can be shown to be an increasing function (under certain assumptions). This is the 
Boltzmann \(H\) theorem \cite{Boltzmann, notcommon}.  
Identifying \(H(t)\) with the disorder of the system, it may seem that the Boltzmann \(H\) theorem is 
against the hypothesis of preservation of disorder. However the \(H\) theorem is true in the limit of 
``molecular chaos'' \cite{chaos}, which means that in a volume element \(d\vec{r}\) around the position \(\vec{r}\), 
the probability of finding two particles with momenta \(\vec{p}_1\) and \(\vec{p}_2\) are independent, at 
any instant of time \(t\), so that the probability of finding them simultaneously 
is the product \(f(\vec{r}, \vec{p}_1, t) f(\vec{r}, \vec{p}_2, t)\) of the individual probabilities. 
It is known that under such an assumption of factorization, an otherwise closed 
system turns into an open system, leading to irreversibility in the process (see e.g. \cite{Calzetta}).

Our considerations are  however for reversible processes, and 
so the \(H\) theorem 
is not relevant for our purposes.
Remaining still within the 
realm of classical mechanics, the Liouville density \(D\) of a system of 
\(N\) (classical) particles in the \(6N\)-dimensional phase space can be 
used to define the disorder of  the \(N\) particle gas
as the Gibbs 
entropy 
\[
H_G(t) = -\int d\Gamma  D \log_2 D,
\]
where \(d\Gamma\) is an element of the phase space. Using Liouville's theorem 
(stating that \(D\) is a constant of motion) \cite{Liouville},
it is straightforward to show that the Gibbs entropy is a constant of (classical) motion.

Similar considerations hold in the quantum domain. It is known \cite{Partovi} (see also \cite{Peres_abar})
 that a quantum system approaches its equilibrium state through multiple collisions with 
other systems in equilibrium. In the process of reaching this 
equilibrium, the von Neumann entropy of the system increases, compatible with
the average energy of the system. In the absence of any such interactions with other external systems, 
the von Neumann entropy of the system remains constant, as a consequence of the unitarity of its evolution.
For an interacting system, the transformation of the system along with the 
heat bath is again
unitary, and hence the entropy of the system plus 
the heat bath is again preserved. However,
when we consider the 
system only, we see an increase of entropy. Such a system is open, and leads, quite 
generally, to irreversibility.

\section{}
\label{sec:appB}

In this appendix we close the gap in the argument concerning the
probabilistic measure of entanglement. We have to 
deal with the case when for the $\lambda_1\mu_1$ and $\lambda_n\mu_m$ of Eq.
(\ref{nonmax}), the inequality 
$\lambda_1\mu_1 \geq 1 \geq \lambda_n \mu_m$
holds. 
Due to continuity of the lengths of the vectors, there  exists a $c_2$ such that
the the normalized vector $|\psi(c_2)\rangle$ (of the form in Eq.
(\ref{vorl})) is mapped by $L$ to a normalized vector (in the form in
Eq. (\ref{nonmax})), i.e.
\begin{align}
|\psi (c_2)\rangle:=c_2 |a_1b_1\rangle +\sqrt{1-c_2^2} |a_nb_m\rangle
\nonumber \\ 
\xrightarrow{L}  
l|x_1y_1\rangle + \sqrt{1-l^2} |x_ny_m\rangle.
\end{align}
Due to preservation of entanglement by $L$ 
the Schmidt values of the two vectors have to be preserved,
and we have that either $c_2^2 \lambda_1^2\mu_1^2=c_2^2$ and $(1-c_2^2)
\lambda_n \mu_m=(1-c_2^2)$, which immediately results in 
$\lambda_1 \mu_1=1=\lambda_n\mu_m$, or 
$c_2^2 \lambda_1^2\mu_1^2=1-c_2^2$ and $(1-c_2^2)\lambda_n \mu_m = c_2^2$.
The last two equations can be combined to 
\begin{equation}
\lambda_1\mu_1\lambda_n \mu_m =1.
\end{equation}
Now, since $L$ is of full rank, every vector is in the range of $L$.
So there exists a $c_3$ such that
\begin{align}
|\psi (c_3)\rangle:=c_3 |a_1b_1\rangle +\sqrt{1-c_3^2} |a_nb_m\rangle
\nonumber \\
\xrightarrow{L}  
\sqrt{E(c_3)} \frac{1}{\sqrt{2}} (|x_1y_1\rangle +|x_ny_m\rangle),
\end{align}
where $E(c_3)$ is the entanglement of $|\psi (c_3)\rangle$, which only
  depends on  
$c_3$. 
We therefore have the requirement 
\begin{align}
\lambda_1\mu_1= \sqrt{\frac{E(c_3)}{2 c_3^2}}, & & \lambda_n \mu_m =
\sqrt{\frac{E(c_3)} {2 (1-c_3^2)}}.
\end{align} 
Using the fact that 
the product of the two terms has to equal 1, as shown before, we have 
that
\begin{equation}
\frac{E(c_3)}{\sqrt{c_3^2(1-c_3^2)}}=2,
\end{equation}
which can easily be seen to have the only positive solution 
$c_3=\frac{1}{\sqrt{2}}$. But this corresponds to the case where it is a 
maximally entangled state that is 
taken to a maximally entangled state, and $\lambda_1\mu_1=1$ as well as 
$\lambda_n\mu_m=1$. So one can take $\lambda_i=1$ and $\mu_j=1$, which shows 
by Eq. (\ref{loc-unit}) that $L_A$ and $L_B$ are actually unitary maps.
So $L$ is the 
product of two local unitaries, and not the multiple thereof, like in
the 
case of the renormalized measure of entanglement.  


\begin{thebibliography}{10}

\bibitem{somebody} J.S. Bell, \emph{Speakable and Unspeakable in Quantum Mechanics} (Cambridge University
Press, Cambridge, 1987); L.E: Ballentine, editor, \emph{Foundations of Quantum Mechanics since the Bell Inequalities}, Amer. Assoc, Phys. Teachers, 
College Park (1988), and references therein.

\bibitem{EPR} A. Einstein, B. Podolsky, and N. Rosen, Phys. Rev. \textbf{47}, 777 (1935).

\bibitem{Bell} J.S. Bell, Physics \textbf{1}, 195 (1964). J.S. Bell, Rev. Mod. 
Phys. \textbf{38}, 447 (1966). 

\bibitem{loopholes} E. Santos, Phys. Rev. Lett. \textbf{66}, 1388 (1991);  E. Santos, Phys. Rev. A \textbf{46}, 3646 (1992).

\bibitem{others} 
P.M. Pearle, Phys. Rev. D \textbf{2}, 1418 (1970);
J.F. Clauser and M.A. Horne, Phys. Rev. D \textbf{10}, 526 (1974);
P.G. Kwiat, P.H. Eberhard, A.M. Steinberg, and R.Y. Chiao, Phys. Rev. A \textbf{49}, 3209 (1994);
N. Gisin and B. Gisin, Phys. Lett. A \textbf{260}, 323 (1999);
S. Massar, S. Pironio, J. Roland, and B. Gisin, Phys. Rev. A \textbf{66}, 052112 (2002);
    R. Garcia-Patron, J. Fiur\'asek, N.J. Cerf, J. Wenger, R. Tualle-Brouri, and Ph. Grangier,  Phys. Rev. Lett. \textbf{93}, 130409 (2004).






\bibitem{CT} C. Cohen-Tannoudji, B. Diu, and F. Lalo{\" e}, \emph{Quantum 
Mechanics} (John Wiley and Sons, NY, and Hermann, Paris, 1977).


\bibitem{Gleason}A.M. Gleason, J. Math. Mech. \textbf{6}, 885 (1957).


\bibitem{KS} S. Kochen and E.P. Specker, J. Math. Mech. \textbf{17}, 59 (1967).

\bibitem{cloning}W.K. Wootters and W.H. Zurek, Nature \textbf{299}, 802 (1982);
D. Dieks, Phys. Lett. A \textbf{92}, 271 (1982); H.P. Yuen, Phys. Lett. A \textbf{113}, 405 (1986). 

\bibitem{GHZ123}D.M. Greenberger, M.A. Horne, and A. Zeilinger, in \emph{Bell's
 Theorem, Quantum Theory, and Conceptions of the Universe}, edited by M. Kafatos
 (Kluwer Academic,
 Dordrecht, The Netherlands, 1989).



\bibitem{Partovi} M.H. Partovi, Phys. Lett. A \textbf{137}, 440 (1989).


\bibitem{Hardy123} L. Hardy,
Phys. Rev. Lett. \textbf{71}, 1665 (1993).


\bibitem{seijey1} R. Olkiewicz, Comm. Math. Phys. \textbf{208}, 245 (1999).

\bibitem{deleting} A.K. Pati and S.L. Braunstein, Nature \textbf{404}, 131
(2000); \emph{ibid.}, quant-ph/0007121. 

\bibitem{seijey56} R. Olkiewicz, Ann. Phys. \textbf{286}, 10 (2000).

\bibitem{Hardy} L. Hardy, quant-ph/0101012.

\bibitem{Dietz}K. Dietz, J. Phys. A - Math. Gen. \textbf{35}, 10573 (2002);
\emph{ibid.}, \textbf{36}, 5595 (2003); 
\emph{ibid.}, \textbf{36}, L45 (2003).


\bibitem{Fuchs} C.M. Caves, C.A. Fuchs, K. Manne, and J.M. Renes, 
Found. Phys. \textbf{34}, 193 (2004), and references therein.

\bibitem{nonlinear}
A. Peres, \emph{Quantum Theory: Concepts and Methods} (Kluwer, Dordrecht, 1995),  
p. 278.


\bibitem{ulto} An antiunitary operator can be always written in the form 
\(U {\cal C}\), where \(U\) is a unitary operator and \({\cal C}\) is 
complex conjugation in the choosen basis.


\bibitem{Huang} K. Huang, \emph{Statistical Mechanics}, (John Wiley \& Sons, New York, 1987).


\bibitem{Welsh} D. Welsh, \emph{Codes and Cryptography}, (Clarendon Press, Oxford, 1989).

\bibitem{Wehrl} A. Wehrl,
Rev. Mod. Phys. \textbf{50}, 221 (1978).


\bibitem{vN1} J. von Neumann, G{\"o}tt. Nachr.,  271 (1927).

\bibitem{Wigner} E.P. Wigner, \emph{Group Theory}, (Academic Press, New York, 1959).

\bibitem{Emch} G. Emch and C. Piron, J. Math. Phys. \textbf{4}, 469 (1963).

\bibitem{Gisin} N. Gisin, Am. J. Phys. \textbf{61}, 86 (1993).

\bibitem{entanmeasure} There are many ways in which one can quantify entanglement. A partial list includes
\cite{onek}.

\bibitem{onek}
C.H. Bennett, D.P. DiVincenzo, J.A. Smolin, and W.K. Wootters,
Phys. Rev. A \textbf{54}, 3824 (1996); 
V. Vedral,  M.B. Plenio,  M.A. Rippin, and P.L. Knight,
Phys. Rev. Lett \textbf{78}, 2275 (1997);
%
D.P. DiVincenzo,
C.A. Fuchs, H. Mabuchi, J.A. Smolin, A. Thapliyal, A.  Uhlmann,
  quant-ph/9803033;
 T. Laustsen, F. Verstraete, and S.J. van Enk,
QIC \textbf{3}, 64 (2003);
%
M.A. Nielsen,
Phys. Rev. Lett.
\textbf{83}, 436 (1999);
G. Vidal,
J. Mod. Opt. \textbf{47}, 355 (2000);
D. Jonathan and  M.B. Plenio, Phys. Rev. Lett. \textbf{83} 
 1455 (1999); 
 %
 M. Horodecki, A. Sen(De), and U. Sen, Phys. Rev. A \textbf{70}, 052326 (2004); 
 %
M. Horodecki, QIC \textbf{1}, 7 (2001).


\bibitem{horo99} C.H. Bennett, H.J. Bernstein, S. Popescu, and B. Schumacher, Phys. Rev. A \textbf{53}, 2046 (1996);
S. Popescu and D. Rohrlich, Phys. Rev. A \textbf{56}, 3219 (1997);
G. Vidal, J. Mod. Opt. \textbf{47}, 355 (2000);
M. Horodecki, P. Horodecki, R. Horodecki, Phys. Rev. Lett. \textbf{84}, 2014 (2000).

\bibitem{Schmidt} E. Schmidt, Math. Ann. {\bf 63}, 433ff (1907).


\bibitem{Boltzmann} See e.g. Ref. \cite{Huang}, p. 73.


\bibitem{notcommon} Note that the usual statement of the Boltzmann \(H\) theorem 
is with the function \(H(t)\), without the 
minus sign.

\bibitem{chaos} See e.g. Ref. \cite{Huang}, p. 62.

\bibitem{Calzetta}  E.A. Calzetta and  B.L. Hu,  Phys. Rev. D \textbf{37}, 2878 (1988); 
\emph{ibid.} \textbf{68}  065027 (2003), and references
therein.

\bibitem{Liouville} See e.g. Ref. \cite{Huang}, p. 64.

\bibitem{Peres_abar} Ref. \cite{nonlinear}, p. 267.




\bibitem{thermo} S. Popescu and D. Rohrlich, Phys. Rev. A \textbf{56}, R3319
(1997); M. B. Plenio and V. Vedral, Cont. Phys. \textbf{39}, 431 (1998); M.
Horodecki and R. Horodecki, Phys. Lett. A \textbf{244}, 473 (1998); P.
Horodecki, M. Horodecki and R. Horodecki, Acta Phys. Slovaca \textbf{48}, 141
(1998).
               

\end{thebibliography}
\end{document}